\documentclass[12pt,twocolumns]{iopart}
 
\expandafter\let\csname equation*\endcsname\relax
\expandafter\let\csname endequation*\endcsname\relax

\usepackage{amsmath}
 
 \usepackage{wrapfig} 
\usepackage{hyperref}       
\usepackage{url}           
\usepackage{booktabs}      
\usepackage{amsfonts}        
\usepackage{nicefrac}       
\usepackage{microtype}   
 
\usepackage[T1]{fontenc}

\usepackage{bm}
\usepackage{braket}
\usepackage{graphicx}
 
\usepackage{bbm}

\newtheorem{thm}{Theorem}

\usepackage{mathtools}
\usepackage{xcolor}
\usepackage{caption}
\DeclareMathOperator{\MCZ}{MCZ}

\begin{document}

\title{A Grover-search Based Quantum Learning Scheme for Classification}

\author{Yuxuan Du $^1$, Min-Hsiu Hsieh $^2$, Tongliang Liu$^1$, Dacheng Tao$^1$}
\address{$^1$ UBTECH Sydney Artificial Intelligence Centre and the School of Information Technologies, Faculty of Engineering and Information Technologies, The University of Sydney, Australia}
\address{$^2$ Centre for Quantum Software and Information, Faculty of Engineering and Information Technology, University of Technology Sydney, Australia}

\vspace{10pt}
 
\begin{abstract}
The hybrid quantum-classical learning scheme provides a prominent way to achieve quantum advantages on near-term quantum devices. A concrete example towards this goal is the quantum neural network (QNN), which has been developed to accomplish various supervised learning tasks such as classification and regression. However, there are two central issues that remain obscure when QNN is exploited to accomplish classification tasks. First, a quantum classifier that can well balance the computational cost such as the number of measurements and the learning performance is unexplored. Second, it is unclear whether quantum classifiers can be applied to solve certain problems that outperform their classical counterparts.  Here we devise a Grover-search based quantum learning scheme (GBLS) to address the above two issues.  Notably, most existing QNN-based quantum classifiers can be seamlessly embedded into the proposed scheme. The key insight behind our proposal is reformulating the classification tasks as the search problem.  Numerical simulations  exhibit that GBLS can achieve comparable performance with other quantum classifiers under various noise settings, while the required number of measurements is dramatically reduced. We further demonstrate a potential quantum advantage of GBLS over classical classifiers in the measure of query complexity.  Our work provides guidance to  develop advanced quantum classifiers on near-term quantum devices and opens up an avenue to explore potential quantum advantages in various classification tasks.
\end{abstract}

\section{Introduction}

The field of machine learning has achieved remarkable success in computer vision, natural language processing, and data mining \cite{goodfellow2016deep}. Recently, an increasing interest from the physics community to use machine learning methods to solve complicated physics problems, e.g., classifying phases of matter and simulating quantum systems \cite{carrasquilla2017machine,van2017learning,carleo2017solving}, has emerged. Besides the revolutionary influence of machine learning to the physics world, another uprising field that tightly binds machine learning with physics is quantum machine learning whose goal is to solve specific tasks beyond the reach of classical computers \cite{biamonte2017quantum}.  

To better understand how quantum computing facilitates the machine learning tasks,  devising quantum algorithms that have the ability to solve fundamental machine learning problems with quantum advantages is desirable  \cite{biamonte2017quantum}. For example, the proposed quantum linear systems algorithm (a.k.a., HHL algorithm) enables the linear equations to be solved with the exponential speedup over its classical counterparts \cite{harrow2009quantum}. By employing HHL algorithm as the subroutine, many quantum machine learning algorithms with exponential quantum speedup have been proposed, e.g., the quantum principal component analysis \cite{lloyd2014quantum}, quantum singular value decomposition \cite{wang2017quantum}, quantum non-negative matrix factorization \cite{du2018quantum}, and the quantum regression  \cite{rebentrost2018quantum}. However, those proposed quantum algorithms that possess fabulous quantum advantages can only be executed on a fault-tolerant quantum computer by using the quantum random access memory \cite{harrow2009quantum}, which is still a rather distant dream.  

When approaching the noisy intermediate-scale quantum (NISQ) era, it is intrigued to explore whether there exists any quantum algorithm that can not only solve fundamental learning problems with promised quantum advantages but can also be efficiently implemented on near-term quantum devices   \cite{preskill2018quantum}. To achieve this goal,  one of the most likely solutions is the quantum neural network (QNN), which is also called as \textit{variational quantum algorithms} \cite{benedetti2019parameterized,du2020learnability,cerezo2020variational}. Concretely, QNN is composed of a variational quantum circuit to prepare quantum states and a classical controller to perform optimization tasks \cite{du2020learnability,farhi2018classification}. Partial evidence to support this claim is the theoretical result that the probability distribution generated by the variational quantum circuit used in QNN  can not be efficiently simulated by classical computers \cite{arute2019quantum,farhi2016quantum,bremner2010classical}.  Driven by the strong expressive power of quantum circuits and the similar work philosophy between QNN and the classical deep neural network (DNN), its natural to exploit whether QNN can be realized on near-term quantum computers to accomplish certain machine learning tasks with better performance over classical learning algorithms. 

A central application of QNN, analogous to DNN, is tackling classification tasks \cite{goodfellow2016deep}. Many real-world problems can be categorized into the classifying scenario, e.g., the recognization of hand-written digits, the characterization of different creatures, and the discrimination of quantum states. For binary classification, given a dataset 
\begin{equation}\label{eqn:def_data}
	\hat{\mathcal{D}}=\{(\bm{x}_i,y_i)\}_{i=0}^{N-1}\in(\mathbb{R}^{N\times M}, \{0,1\}^{N})~,
\end{equation}
with $N$ examples and $M$ features in each example, QNN aims to learn a decision rule $f_{\bm{\theta}}(\cdot)$ that correctly predicts the label of the given dataset $\mathcal{\hat{D}}$, i.e., 
\begin{equation}\label{eqn:classify_aim}
	\min_{\bm{\theta}} \sum_{i=0}^{N-1} \mathbbm{1}_{y_i\neq f_{\bm{\theta}}(\bm{x}_i)}~,
\end{equation}
where $\bm{\theta}$ refers to the trainable parameters and $\mathbbm{1}_{z}$ is the indicator function that takes the value $1$ if the condition $z$ is satisfied and zero otherwise.  Recently, QNNs with varied quantum circuit architectures and optimization methods have been proposed to accomplish the aforementioned classification tasks. In particular, the references  \cite{schuld2017implementing,schuld2020circuit,wilson2018quantum} have devised the amplitude encoding based QNN to classify the Iris dataset and the hand-written digits image dataset; the references  \cite{mitarai2018quantum,havlivcek2019supervised,schuld2019quantum} have developed the kernel-based QNN to accomplish the synthetic datasets; and the references \cite{cong2019quantum} have proposed the convolution based QNN to tackle quantum state discrimination tasks. When no confusion can arise, we use the \textit{quantum classifier} in the rest of the study to specify QNNs that are used to accomplish classification tasks defined in Eqn.~(\ref{eqn:classify_aim}). 

\begin{figure*}
	\centering
	\includegraphics[width=0.98\textwidth]{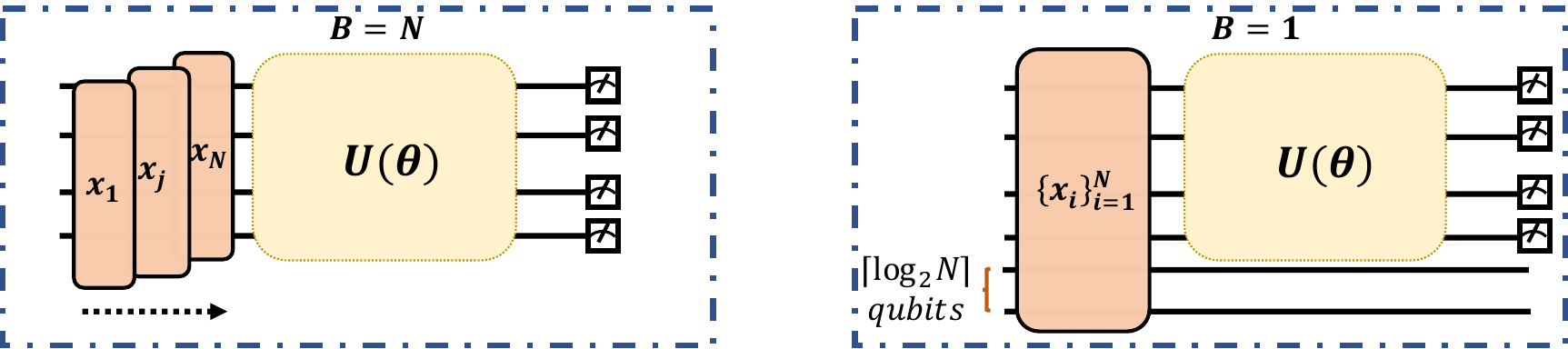}
	\caption{\small{\textbf{The protocol of the batch gradient descent method}. The left panel corresponds to the setting as $B=N$, where the $N$ training examples $\{\bm{x}_i\}_{i=1}^N$ are iteratively fed into the variational quantum circuits to acquire the gradients that estimate $\{\nabla\mathcal{L}(\bm{\theta}^{(t)}, \bm{x}_i)\}_{i=1}^N$. The right panel exhibits the implementation of the quantum classifier when $B=1$. Specifically,  a superposition state $\ket{\phi(\bm{x})}=\frac{1}{\sqrt{N}}\sum_{i=1}^N\ket{h(\bm{x}_i)}_F\ket{i}_I$ is prepared, where $h(\cdot)$ corresponds to the employed encoding method and the subscripts `I' and `F' refer to the index and feature registers, respectively.  Given access to $\ket{\phi(\bm{x})}$, the trainable quantum circuit $U_L(\bm{\theta})$ is employed to interact with its feature register subscripted with $F$.   } }
	\label{fig:BGD}
\end{figure*}

Despite the promising heuristic results mentioned above, very few studies have theoretically explored the power  of quantum classifiers. A noticeable theoretical result about quantum classifiers is the trade-off between the computational cost (i.e., the number of measurements) and the training performance indicated by \cite{du2020learnability}. Denote $\mathcal{L}(\bm{\theta}^{(t)},\bm{z})$ as the loss function employed in quantum classifiers, where $\bm{\theta}^{(t)}$ refers to the trainable parameters at the $t$-th iteration and $\bm{z}=\{\bm{z}_j\}_{j=1}^N$ is the given dataset with in total $N$ samples. As shown in  Figure \ref{fig:BGD}, when the \textit{batch gradient descent} method is employed to optimize the loss function $\mathcal{L}$, the updating rule of the trainable parameters follows 
\begin{equation}\label{eqn:opt_batch}
	\bm{\theta}^{(t+1)} = \bm{\theta}^{(t)} - \frac{\eta}{B}\sum_{i=1}^B\nabla\mathcal{L}(\bm{\theta}^{(t)}, \mathcal{B}_i),  
\end{equation}
where $\eta$ is the learning rate, $\mathcal{B}_i$ refers to the $i$-th batch with $\cup_{i=1}^B \mathcal{B}_i = \bm{z}$ and $\mathcal{B}_i \cap \mathcal{B}_j=\emptyset$, and $B$ denotes the number of batches.  Define 
\begin{equation}
R_1=\mathbb{E}[ \|\nabla_{\bm{\theta}}\mathcal{L}(\bm{\theta}^{(t)})\|^2].	
\end{equation}
as the utility measure that evaluates the distance between the optimized result and the stationary point in the optimization landscape.  The following theorem summarizes the utility bound $R_1$ of quantum classifiers.      
\begin{thm}[Modified from Theorem 1 of \cite{du2020learnability}.] \label{thm:informal_utl_QNNQAE_DP}
Quantum classifiers under the depolarization noise setting output $\bm{\theta}^{(T)}\in\mathbb{R}^d$ after $T$ iterations with the utility bound  
\[
R_1   \leq  \tilde{O}\left(poly\left(\frac{d}{T(1-p)^{L_Q}}, \frac{d}{BM(1-p)^{L_Q}}, \frac{d}{(1-p)^{L_Q}} \right) \right),
\] 
where $M$ is the number of measurements to estimate the quantum expectation value, $L_Q$ is the circuit depth of variational quantum circuits, $p$ is the rate of the depolarization noise, and $B$ is the number of batches. 
\end{thm} 
The result of Theorem \ref{thm:informal_utl_QNNQAE_DP} indicates that a larger number of batches $B$ ensures a better utility bound $R_1$, while the price to pay is increasing the total number of measurements. For example, when $B=N$, we have $\mathcal{B}_i=\bm{z}_i$ for $\forall i \in [N]$ and each sample $\bm{z}_j$ is  sequentially fed into variational quantum circuits to acquire $\nabla \bar{\mathcal{L}}(\bm{\theta}, \bm{z}_i)$ that estimates $\nabla \mathcal{L}(\bm{\theta}, \bm{z}_i)$. Once the set $\{\nabla \mathcal{L}(\bm{\theta}, \bm{z}_i)\}_{i=1}^N$ is collected, the gradients $\nabla \mathcal{L}(\bm{\theta}, \bm{z})$ can be estimated by $ \frac{1}{N}\sum_{i=1}^N \nabla \bar{\mathcal{L}}(\bm{\theta}, \bm{z}_i)$. Suppose that the required number of measurements to estimate the derivative of the $j$-th parameter $\bm{\theta}_j$, i.e., $\nabla_j \mathcal{L}(\bm{\theta}, \bm{z}_i)=\frac{\partial \mathcal{L}(\bm{\theta}, \bm{z}_i)}{\partial \bm{\theta}_j}$, is $M$, then the total number of measurements to acquire $\frac{1}{N}\sum_{i=1}^N \nabla_j \bar{\mathcal{L}}(\bm{\theta}, \bm{z}_i)$ is $NM$. Therefore, the estimation of $\nabla \mathcal{L}(\bm{\theta}, \bm{z})$, which includes $d$ parameters, requires $NMd$ measurements.  Such a cost becomes unaffordable for large $N$.  However, the trade-off between the utility $R_1$ and the computational efficiency caused by the varied number of batches $B$ is not considered in previous quantum classifiers, where most of them only focused on the setting $B=N$. How to design a quantum classifier that can attain a good utility $R_1$ with a low computational cost is unknown.

Another theoretical issue towards quantum classifiers is that none of the previous results have explored their potential advantages compared with  classical counterparts. This questions the necessity of employing quantum classifiers because no benefit can be offered. Under the above observations, it is highly desirable to develop a quantum classifier that can not only achieve a good utility $R_1$ using a low computational cost, but can also possess certain quantum advantages compared with classical classifiers. 

Here we devise a Grover-search based learning scheme (GBLS) to address the above two issues under the NISQ setting. Our proposal has the following advantages. First, GBLS is a flexible and effective learning scheme, which enables the optimization of different quantum classifiers with a varied number of batches $B$. Note that the choice of the encoding methods and the variational ansatz used in GBLS is very flexible, which covers a wide range of the proposed quantum classifiers  \cite{schuld2020circuit,wilson2018quantum,mitarai2018quantum,havlivcek2019supervised,schuld2019quantum}. Moreover, the Grover-search based machinery is only required in the training process, and the prediction of the new input is completed by only using the optimized variational quantum circuits, which ensures its efficacy. Second, we prove that the query complexity can be quadratically reduced over its classical counterparts in the optimal setting (see Theorem \ref{thm1}) when it is applied to accomplish specific binary classification tasks.  Last, numerical simulation results demonstrate that GBLS can well accomplish binary classification tasks even when the system noise and the finite number of quantum measurements are considered (see Section \ref{sec:numerical_sim}). Notably, the required number of measurements of GBLS is dramatically less than other advanced quantum classifiers \cite{havlivcek2019supervised,mitarai2018quantum,schuld2019quantum} with competitive performance (see Table \ref{tab:my-table}). In other words, GBLS is a powerful protocol that allows quantum classifiers to achieve a good utility bound $R_1$ with a low computational cost.

\begin{figure}[!ht]
 \centering
\includegraphics[width=0.695\textwidth]{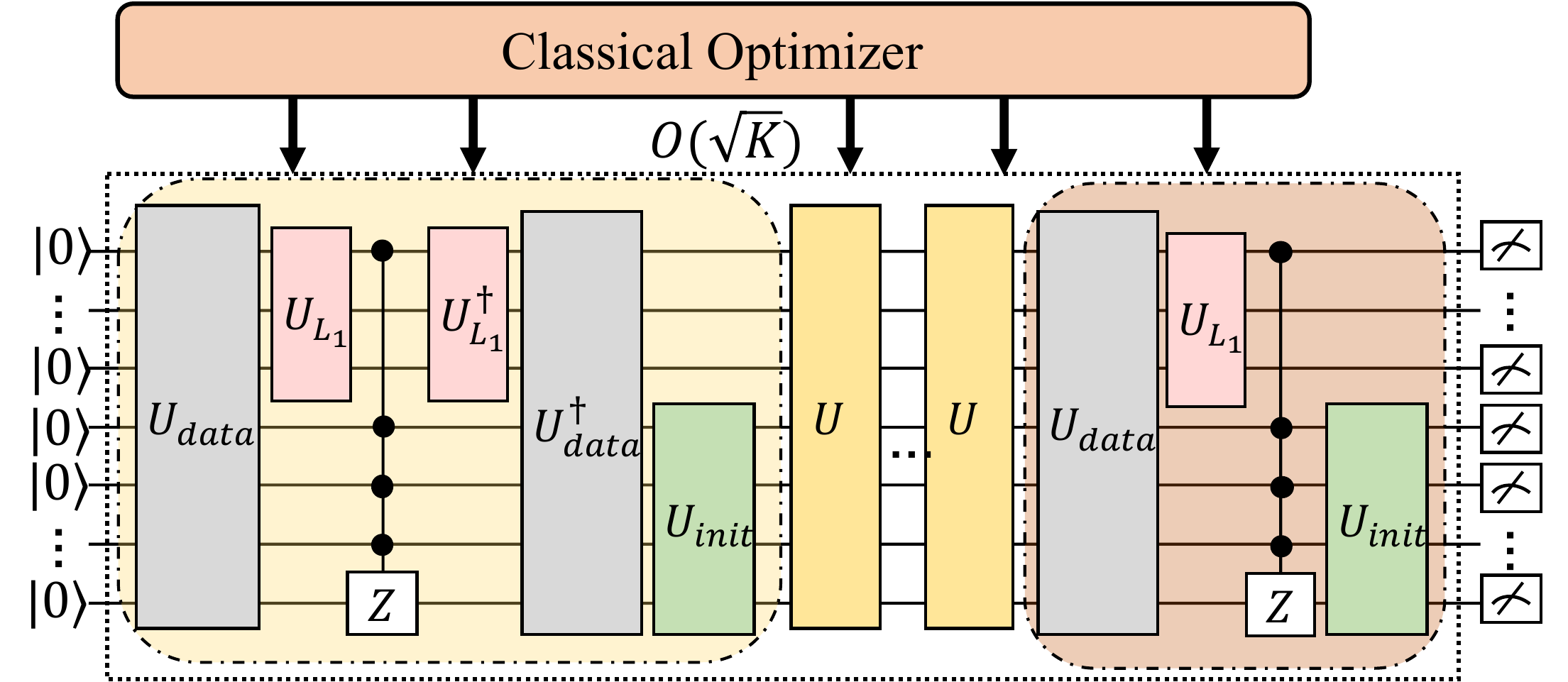}
\caption{The paradigm of GBLS. $U$ defined in Eqn.~(\ref{def:U}) is composed of unitary operators (i.e., $U_{data}$, $U_{L_1}$, MCZ, and $U_{init}$) highlighted by the shadowed yellow region.  The last cycle employs the unitary operation $U_E$ defined in Eqn.~(\ref{eqn:U_E}), highlighted by the brown region. The qubits interacted with $U_{L_1}$ (or $U_{init}$) form the feature (or data) register $\mathcal{R}_F$(or $\mathcal{R}_I$).}
\label{fig:GBLS}  
\end{figure}

The central concept in GBLS is reformulating the classification tasks as the search problem. Note that although the advantage held by the quantum Grover-search algorithm is evident, how to transform the classification task into the search problem is \textit{inconclusive}. Such a reformulation is the main technical contribution in this study. Recall that Grover-search \cite{grover1996fast}  identifies the target element $i^*$ in a database of size $K$ by iteratively applying a predefined oracle $U_f=\mathbb{I}-2\ket{i^*}\bra{i^*}$ and a diffusion operator $U_{init}=2\ket{\varphi}\bra{\varphi} - \mathbb{I}$ with $\ket{\varphi}=\frac{1}{\sqrt{K}}\sum_i \ket{i}$ to the input state.  GBLS, as shown in Figure \ref{fig:GBLS}, employs a specified variational quantum circuit $U_{L_1}$ and a multiple controlled qubits gate along the $Z$ axis ($\MCZ$) to replace the oracle $U_f$. In particular, the variational quantum circuit conditionally flips a flag qubit (i.e., the black dot behind $U_{L_1}$ highlighted by the pink region) depending on the training data. The flag qubit is then employed as a part of MCZ gate to guide a Grover-like search algorithm to identify the index of the specified example, i.e., the status of the flag qubit such as `0' or `1' determines the successful probability to identify the target index. Through optimizing the trainable parameters of the variational quantum circuits $U_{L_1}$, GBLS aims to maximize the successful probability to sample the target index when the corresponding training example is positive; otherwise, GBLS minimizes the successful probability of sampling the target index. The inherited property from the Grover-search algorithm allows our proposal to achieve an advantage in terms of query complexity when the binary classification task involves the searching constraint (see Section \ref{subsec:adv_GBLS} for details).   Besides the computational merit, GBLS is insensitive to noise, guaranteed by the fact that combining a variational learning approach with Grover-search can preserve a high probability of success in finding the solution under the NISQ setting \cite{morales2018variational}.

\section{Grover-search based learning scheme}

The outline of this section is as follows. In Subsection \ref{subsec:implement}, we first elaborate on the implementation details of the proposed Grover-search based learning scheme (GLBS) as depicted in Figure \ref{fig:GBLS}. We then explain how to use the trained GLBS to predict the given new input with $O(1)$ query complexity in Subsection \ref{subsec:predict}. We last explain how GBLS can solve certain learning problems with potential advantages in Subsection \ref{subsec:adv_GBLS}.

\subsection{Implementation}\label{subsec:implement}
 In the preprocessing stage, GBLS employs the dataset $\hat{\mathcal{D}}$ defined in Eqn.~(\ref{eqn:def_data}) to  construct an \textit{extended} dataset $\mathcal{D}$. Compared with the original dataset $\hat{\mathcal{D}}$, the cardinality of each training example in $\mathcal{D}$ is enlarged to $K$. For the purpose of applying the Grover-search algorithm to locate the target index $i^*=K-1$, the construction rule for the $k$-th extended training example $\mathcal{D}_k$ for all $k\in[N]$ is as follows. The mathematical representation of $\mathcal{D}_k$  is 
 \begin{equation}\label{eqn:ext_train_exp}
  \mathcal{D}_k=[(\bm{x}_{k}^{(0)}, y_k^{(0)}), (\bm{x}_{k}^{(1)}, y_k^{(1}),...,(\bm{x}_{k}^{(K-1)}, y_k^{(K-1)})]~.	
 \end{equation}
 The last pair in $\mathcal{D}_k$  corresponds to the $k$-th example of $\hat{\mathcal{D}}$, i.e., $(\bm{x}_{k}^{(K-1)}, y_k^{(K-1)})= (\bm{x}_{k}, y_{k})$. The first $K-1$ pairs $\{(\bm{x}_{k}^{(i)}, y_k^{(i)})\}_{i=0}^{K-2}$ in $\mathcal{D}_k$ are uniformly sampled from a subset of $\hat{\mathcal{D}}$, where all labels of this subset, i.e., $\{y_k^{(i)}\}_{i=1}^{K-2}$, are opposite to $y_k$. Note that the construction of the subset is efficient. Since $y_k\in\{0, 1\}$, we can construct two subsets $\hat{\mathcal{D}}^{(0)}$ and $\hat{\mathcal{D}}^{(1)}$ that only contains examples of $\hat{D}$ with  label `0' and label `1', respectively, where $\hat{\mathcal{D}}^{(0)}\cup \hat{\mathcal{D}}^{(1)}= \hat{\mathcal{D}}$. When $y_k=0$, the first $K-1$ pairs are sampled from $\hat{\mathcal{D}}^{(1)}$; otherwise, when $y_k=1$, the first $K-1$ pairs are sampled from $\hat{\mathcal{D}}^{(0)}$.   
 
 \begin{wrapfigure}{r}{0.45\textwidth}
  \begin{center}
    \includegraphics[width=0.36\textwidth]{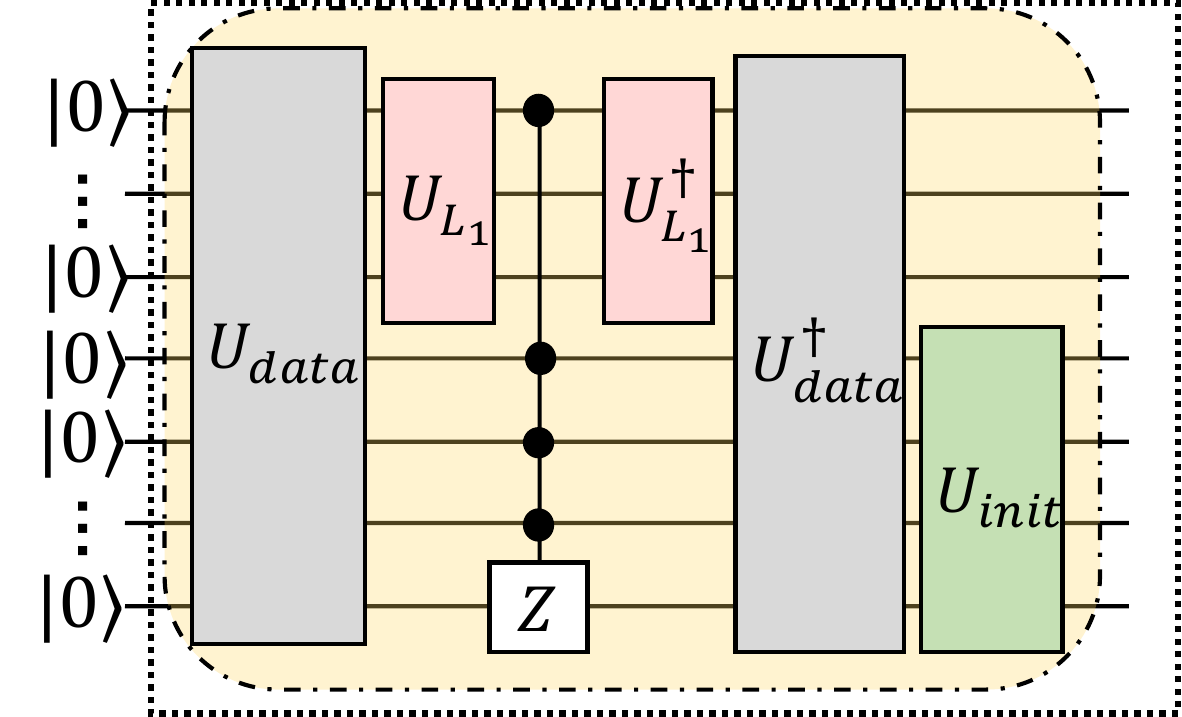}
  \end{center}
  \caption{\small{\textbf{The circuit implementation of the oracle $U$ in Eqn.~(\ref{def:U})}.}}
  \label{fig:GBLS_U}
\end{wrapfigure}

As aforementioned, different quantum classifiers exploit different  methods to encode $\mathcal{D}_k$ into the quantum states \cite{Benedetti_2019}. For ease of notation, we denote the  quantum state corresponding to the $k$-th example $\mathcal{D}_k$  as 
\begin{equation}\label{eqn:encod_0}
U_{data}\ket{\bm{0}}:=\ket{\Phi^k}_{F,I}=\frac{1}{\sqrt{K}}\sum_{i=0}^{K-1}\ket{h(\bm{x}_{i})}_F\ket{i}_I~, 
\end{equation}
 where $h(\cdot)$ is an encoding operation (a possible encoding method is discussed in Section \ref{sec:numerical_sim}),
and the subscripts `$F$' and `$I$' refer to the feature register $\mathcal{R}_F$ with $N_F$ qubits and  the index register $\mathcal{R}_I$ with $N_I$ qubits, respectively.

We now move on to explain the training procedure of GBLS. Recall that the reference \cite{morales2018variational} points out that combining a variational learning approach with Grover-search algorithm produces an additional quantum advantage than conventional Grover’s algorithm such that the target solution can be located with a higher success probability.  A similar idea is used in GBLS. Namely, the employed variational quantum circuits $U_{L_1}$ aim to learn a hyperplane that separates the last pair in $\mathcal{D}_k$ with its first $K-1$ pairs. Denote $U_{L_1}=\prod_{l=1}^{L}U(\bm{\theta}^l)$,  where each layer $U(\bm{\theta}^l)$ contains $O(poly(N_F))$ parameterized single qubit gates and at most $O(poly(N_F))$ fixed two-qubit gates with the identical layouts. In the \textit{optimal} situation, given the initial state $\ket{\Phi^k}_{F,I}$ in Eqn.~(\ref{eqn:encod_0}), applying $U_{L_1}=\prod_{l=1}^{L}U(\bm{\theta}^l)$ to the feature register $\mathcal{R}_F$ yields the following target state: 
\begin{enumerate}
	\item If the \textit{last pair} of the input example $\mathcal{D}_k$ refers to the label  $y_k=0$,  the target state is 
\begin{equation}\label{eqn:lab_0_case}
	(U_{L_1}\otimes \mathbb{I}) \ket{\Phi^k(y_k=0)}_{F,I}= \frac{1}{\sqrt{K}}\sum_{i=0}^{K-1}\ket{\psi_i^{(0)}}_F\ket{i}_I~;
\end{equation}
	\item Otherwise, when the \textit{last pair} of the input example $\mathcal{D}_k$ refers to $y_k=1$, the target state is  \begin{equation}\label{eqn:lab_1_case}
	(U_{L_1}\otimes \mathbb{I}) \ket{\Phi^k(y_k=1)}_{F,I}= \frac{1}{\sqrt{K}}\sum_{i=0}^{K-1}\ket{\psi_i^{(1)}}_F\ket{i}_I~.
\end{equation}
\end{enumerate}

 \begin{wrapfigure}{r}{0.45\textwidth}
  \begin{center}
    \includegraphics[width=0.3\textwidth]{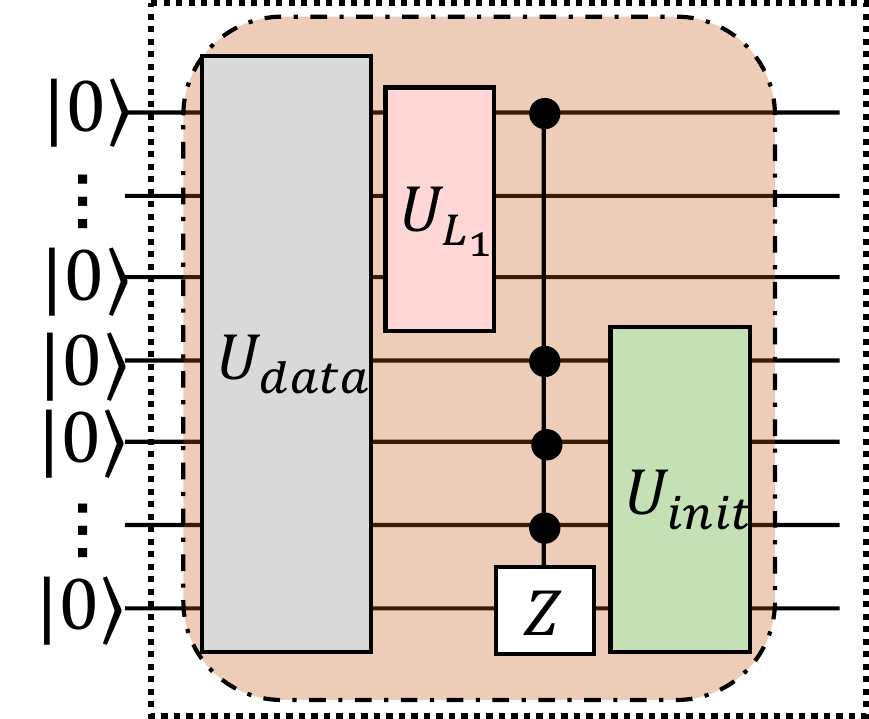}
  \end{center}
  \caption{\small{\textbf{The circuit implementation of the oracle $U_E$ in Eqn.~(\ref{eqn:U_E})}.}}
  \label{fig:GBLS_unit_UE}
\end{wrapfigure}

\noindent We denote $\ket{\psi_i^{(0)}}_{F}$ (resp.~$\ket{\psi_i^{(1)}}_{F}$) as the first qubit of the quantum state  in the feature register $\mathcal{R_F}$ being $\ket{0}$ (resp.~$\ket{1}$).  As shown in Figure \ref{fig:GBLS_U}, once the  state $(U_{L_1}\otimes \mathbb{I}_I) \ket{\Phi^k}_{F,I}$ is prepared, GBLS iteratively applies $\MCZ$ gate to the index register controlled by the first qubit of the feature register and the index register,  uses $U_{data}$ and $U_{L_1}$ to uncompute the feature register, and applies the diffusion operator $U_{init}$ to the index register to complete the first cycle. Denote all quantum operations belong to one cycle as $U$, i.e.,
\begin{equation}\label{def:U}
	U:=U_{init}\circ U_{data}^{\dagger}\circ (U_{L_1}\otimes \mathbb{I})^{\dagger} \circ \MCZ  \circ (U_{L_1}\otimes \mathbb{I})\circ U_{data}~.
\end{equation} 
With a slight abuse of notation, we define   $U_{init}=\mathbb{I}_{F}\otimes  (2\ket{\varphi}\bra{\varphi} - \mathbb{I}_I)$  with $\ket{\varphi}=\frac{1}{\sqrt{K}}\sum_i \ket{i}$ in the rest of the paper. GBLS repeatedly applies $U$ to the initial state $\ket{\bm{0}}$ except for the last cycle, where the applied unitary operations are replaced by 
\begin{equation}\label{eqn:U_E}
	U_E:=U_{init}\circ \MCZ  \circ (U_{L_1}\otimes \mathbb{I})\circ U_{data}~,
\end{equation}
as highlighted by the brown shadow in Figure \ref{fig:GBLS_unit_UE}.  Following the conventional Grover-search, GBLS queries $U$ and $U_E$ with in total  $O(\sqrt{K})$ times before taking quantum measurements. This completes the quantum part of GBLS.

We next analyze how the quantum state evolves for the case $y_k=0$ and $y_k=1$,  respectively. For the case of $y_k=0$, applying $U_{L_1}\otimes \mathbb{I}_{I}$ to the input state $\ket{\Phi^k(y_k=0)}_{F,I}$ in Eqn.~(\ref{eqn:encod_0}) will transform this state to $1/\sqrt{K}\sum_{i=0}\ket{\psi_i^{(0)}}_F\ket{i}_I$ as described in Eqn.~(\ref{eqn:lab_0_case}). Since the control qubit in the feature register is $0$, applying $\MCZ$ gate does not  flip the phase of the state. After  uncomputing, the result state yields   $1/\sqrt{K}\sum_{i=0}\ket{\bm{0}}_F\ket{i}_I$. The positive phase for all computational basis $i\in[K-1]$ implies that applying the quantum operation $U_{init}\circ U_{data}^{\dagger}\circ (U_{L_1}\otimes \mathbb{I}_I)^{\dagger}$ does not change the state as well, i.e., 
\begin{align}\label{eqn:opt_lab0}
	&\left(\mathbb{I}_{F}\otimes  (2\ket{\varphi}\bra{\varphi} - \mathbb{I}_I)\right)\frac{1}{\sqrt{K}}\sum_{i=0}^{K-1}\ket{\bm{0}}_F\ket{i}_I =  \frac{1}{\sqrt{K}}\sum_{i=0}^{K-1}\ket{\bm{0}}_F\ket{i}_I~.
\end{align}  
 In other words, when we measure the index register of the output state, the probability to sample the computation basis $i$ with $i\in[K-1]$ is uniformly distributed.

 For the case of $y_k=1$,  the input state $\ket{\Phi^k(y_k=1)}_{F,I}$ in Eqn.~(\ref{eqn:encod_0}) will be transformed to  $1/\sqrt{K}\sum_{i=0}\ket{\psi_i^{(1)}}_F\ket{i}_I$ after interacting with unitary  $U_{L_1}\otimes \mathbb{I}_I$, as described in Eqn.~(\ref{eqn:lab_1_case}).  With the control qubit in the feature register being $1$, such a generated quantum state will evolve as Grover-search algorithm does by iteratively applying $\MCZ$, the uncomputation operation $U_{data}^{\dagger}\circ (U_{L_1}\otimes \mathbb{I})^{\dagger}$, and $U_{init}$. Mathematically, the result state after interacting with $\MCZ$ yields 
\begin{align}\label{eqn:opt_lab1}
& \hat{U}_{f}\ket{\Phi^k(y_k=1)}_{F,I} = \cos\gamma \ket{\psi_B^{(0)}}_F\ket{B}_I - \sin\gamma \ket{\psi_{i^*}^{(1)}}_F\ket{i^{*}}_I,
\end{align}
where $\hat{U}_f:=  \MCZ \circ (U_{L_1}\otimes \mathbb{I})$,  $ \cos\gamma=\frac{\sqrt{K-1}}{\sqrt{K}}$, $\ket{B}_I=\frac{1}{\sqrt{K-1}}\sum_{i=0}^{K-2}\ket{i}_I$, and $\ket{i^{*}}_I$ refers to the computational basis $\ket{K-1}$.  Analogous to the $U_f$ in Grover-search, the trainable and data-driven $\hat{U}_f$ used above conditionally flips the phase of the state $\ket{i^{*}}$. Next, the uncomputing operation $U_{data}^{\dagger}\circ (U_{L_1}\otimes \mathbb{I})^{\dagger}$ and the diffusion operator $U_{init}$ are employed to increase the probability of $\ket{i^{*}}_I$. Mathematically, the generated state after the first cycle yields
\begin{equation}
\label{eqn:init}
U\ket{\Phi^k(y_k=1)}_{F,I}= \cos3\gamma \ket{\bm{0}}_F\ket{B}_I + \sin3\gamma \ket{\bm{0}}_F\ket{i^{*}}_I,
\end{equation} 
where $U$ is defined in Eqn.~(\ref{def:U}). The probability of sampling $i^*$ is increased to $\sin^2 3\gamma$, which is in accordance to Grover-search algorithm. This observation leads to the following theorem, whose  proof is given in \ref{appd:thm1}. 
    
\begin{thm}\label{thm1}
For GBLS, under the optimal setting, the probability of sampling the outcome $i^*=K-1$  approaches $1$ asymptotically iff the label of the last entry of $\mathcal{D}_k$ is $y_k=1$.
\end{thm}

We leverage the particular property of GBLS, in which  the output distribution is  varied for different label of input $\mathcal{D}_k$ as shown in Theorem \ref{thm1}, to accomplish the binary classification task.   Concisely, the output state of GBLS, i.e., $U_EU^{O(\sqrt{K})}\ket{\bm{0}}_{F,I}$, corresponding to $y_k=1$  will contain the computational basis $i=K-1$ with probability near to $1$.  By contrast, the output state corresponding to $y_k=0$  will contain all computational bases $i\in[K-1]$ with the equal probability. Driven by this observation and the mechanism of the Grover-search algorithm, the loss function of GBLS is  
\begin{equation}\label{eqn:loss}
	\min_{\bm{\theta}}\mathcal{L}(\bm{\theta}):=\text{sign}(1/2-y_k)\Tr(\Pi \rho(\bm{\theta}))~,
\end{equation}
 where $\text{sign}(\cdot)$ is the sign function, $\Pi=(\ket{1}\bra{1})\otimes \mathbb{I}\otimes(\ket{i^*}\bra{i^*})$ refers to the measurement operator, $\rho(\bm{\theta})=U_EU(\bm{\theta})^{O(\sqrt{K})}\ket{\bm{0}}\bra{\bm{0}}(U_EU(\bm{\theta})^{O(\sqrt{K})})^{\dagger}$ is the generated quantum state, and $U(\bm{\theta})$ is defined in Eqn.~(\ref{def:U}) (for clearness, we use the explicit form $U(\bm{\theta})$ instead of $U$). Intuitively, the minimized $\mathcal{L}(\bm{\theta})$ corresponds to the facts that when $y_k=1$ ($y_k=0$), the success probability to sample $i^*$ as well as attain the first feature qubit to be `$1$' (`$0$') is maximized (minimized). GBLS employs a gradient-based method, i.e., the parameter shift rule \cite{mitarai2018quantum}, to optimize $\bm{\theta}$. Confer  \ref{appd:PQC} for the detail. 

We would like to address that, GBLS can be used to conduct both the linear and nonlinear classification tasks depending on the specified quantum classifiers. For example, when GBLS adopts the proposal \cite{havlivcek2019supervised,schuld2019quantum} to implement $U_{data}$ and $U_{L_1}$, it has capability of classifying nonlinear data. 

\subsection{Prediction}\label{subsec:predict}
\begin{wrapfigure}{r}{0.45\textwidth}
  \begin{center}
    \includegraphics[width=0.45\textwidth]{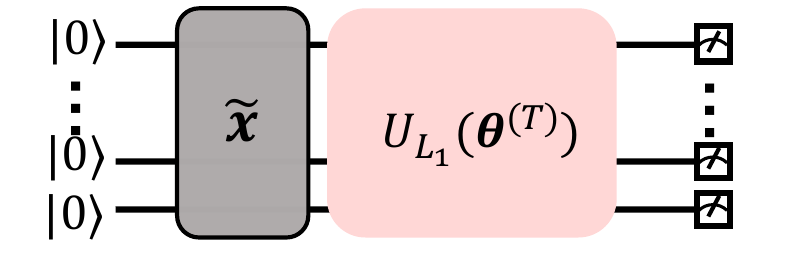}
  \end{center}
  \caption{\small{\textbf{The circuit implementation of GBLS for prediction}. The same encoding method used in the training process is adopted to prepare the state $\ket{h(\bm{\tilde{x}})}$. The trained variational quantum circuit $U(\bm{\theta}^{(T)})$ is applied to $\ket{h(\bm{\tilde{x}})}$ before the measurement. }}
  \label{fig:GBLS_pred}
\end{wrapfigure}

Once the training of GBLS has finished, the trained $U_{L_1}$ can be directly employed to predict the label of the future instances with $O(1)$ query complexity, where the corresponding circuit implementation is shown in Figure \ref{fig:GBLS_pred}. To achieve this, we devise the following prediction method. Denote the new input as $(\tilde{\bm{x}}, \tilde{y})$. We first encode $\tilde{\bm{x}}$ into the quantum state with the identical encoding method used in the training procedure, i.e., $\ket{\tilde{\psi}}_{F} = \ket{h(\tilde{\bm{x}})}$.  Applying the trained $U_{L_1}$ to $\ket{\tilde{\psi}}_{F}$  yields 
\begin{equation}\label{eqn:201}
U_{L_1}\ket{\psi}_{F}= \tilde{\alpha}\ket{\tilde{\psi}^{(0)}}_F + \tilde{\beta}\ket{\tilde{\psi}^{(1)}}_F~,
\end{equation} 
where $|\tilde{\alpha}|^2 + |\tilde{\beta}|^2=1$.  

Denote the probability of the outcome `1' after measuring the first feature qubit of the state in Eq.~(\ref{eqn:201}) as $p_1=|\tilde{\beta}|^2$ and let the threshold be $1/2$. The new input data $\bm{\tilde{x}}$ will be identified as label `0', if $p_1<1/2$; otherwise, it will be given label `$1$'.

\subsection{Potential advantage of GBLS}\label{subsec:adv_GBLS}
Here we  design a binary classification task to explore the potential advantage of GBLS in terms of query complexity. Consider the classification task that requires not only to find a decision rule in Eqn.~(\ref{eqn:classify_aim}) but also to output the index $j$  satisfying a pre-determined black-box function.  Note that the identification of a target index is a common functionality in the context of database searching in the medical system, economy, and online shopping. {For example, given a medical database, it is natural to expect that the trained classifier can predict whether a patient is ill or healthy based on her/his symptoms, and can identify a healthy patient with additional properties, e.g., the gender of the patient is female, which can be modeled by a black box function.}   

The mathematical formulation of this classification task is as follows. Given the data  $\mathcal{D}_k$ in Eqn.~(\ref{eqn:ext_train_exp}), denoted the black box as $q(\cdot)$, the task yields 
\begin{equation}\label{eqn:classi_new}
\left(\min_{\bm{\theta}} \sum_{i=0}^{K-1} \mathbbm{1}_{y_i\neq f_{\bm{\theta}}(\bm{x}_i)}\right) \land
\left(\{j| q(j)=1, \ y_j=1  \}\right), 
\end{equation} 
where the function $q(\cdot)$ is a boolean function with the input set $\{j: \forall y_j \in \mathcal{D}_k, y_j=1  \}$. Taking GBLS implemented in the previous subsections as an example, $q(\cdot)$ has the following form, $\forall j =\{0,\cdots,K-1\}$
\begin{equation}
q(j) =\begin{cases}1, \text{if $j=K-1$;} \\ 0,\text{otherwise.} \end{cases}
\end{equation}
Furthermore, $q(\cdot)$ could be implemented by the MCZ gate, which conditionally flips the phase of the computational basis corresponding to $j^*:=K-1$ if the state is $\ket{\psi_j^{(1)}}_F\ket{j^*}_I$ given in Eqn.~(\ref{eqn:lab_1_case}). In this way, the Grover-like search structure used in GBLS promises that the probability to sample $j^*$ will be maximized. We remark that GBLS can be effectively generalize to implement other forms of $q(\cdot)$ via modifying the MCZ gate. When the size of the dataset loaded by GBLS is $K$, a well-trained  GBLS can locate the target index with $O(\sqrt{K})$ query complexity, guaranteed by the result of Theorem \ref{thm1}. However, given access to the well-trained classifier $f_{\bm{\theta}}(\cdot)$, both classical algorithms and previous quantum classifiers need at least $O(K)$ query complexity to find $j^*$. The reduced query complexity of GBLS implies a potential quantum advantage to accomplish classification tasks.

\begin{figure*}[ht!] 
\centering   
\includegraphics[width=0.95\textwidth]{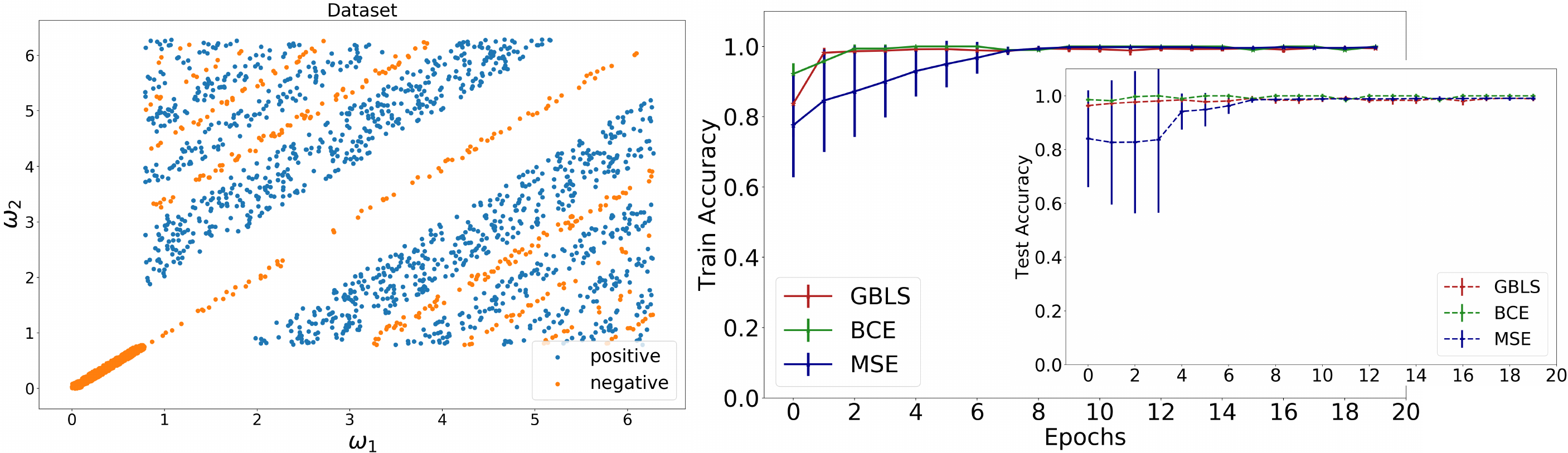}
\caption{The synthetic dataset and  performance of different quantum classifiers under the ideal setting. The left panel illustrates the synthetic dataset used in the numerical simulations. The legend `positive' (or `negative') refers that the label of the data is $1$ (or $0$).  The right panel demonstrates the training and test accuracy of different quantum classifiers. The labels `GBLS', `BCE', `MSE' refer to the proposed GBLS, the quantum kernel classifier with binary cross entropy loss, and the quantum kernel classifier with the mean square error loss ($B=N$)\cite{havlivcek2019supervised,schuld2019quantum}, respectively. The vertical sticks reflect the variance of the test accuracy at each iteration.}
\label{fig:kernel_data}
\end{figure*}

\section{Numerical Experiments}\label{sec:numerical_sim}
We now apply GBLS to classify a nonlinear synthetic dataset $\hat{\mathcal{D}}$ to evaluate its performance. The construction of $\hat{\mathcal{D}}$ follows the proposal \cite{havlivcek2019supervised}.   Consider a synthetic dataset $\hat{\mathcal{D}}=\{\bm{x}_i,y_i\}_{i=0}^{N-1}$ with $N=200$, 
where $\bm{x}_i= (\omega_1^{(i)},\omega_2^{(i)})\in\mathbb{R}^2$, $\omega_1^{(i)}, \omega_2^{(i)} \in(0, 2\pi)$. Let  $g(\cdot)$ be a specific embedding function with $\ket{g(\omega_1^{(i)}, \omega_2^{(i)})}\in\mathbb{C}^4$ for all $i\in \{0,...,N-1\}$.
The label of $\bm{x}_i$ is assigned as $y_i=1$ if 
$$\langle g(\omega_1^{(i)},\omega_2^{(i)})|V^{\dagger}\Pi V|g(\omega_1^{(i)},\omega_2^{(i)}) \rangle \geq 0.5 + \Delta,$$
where $V\in SU(4)$ is a unitary operator, $\Pi=\mathbb{I}\otimes \ket{0}\bra{0}$ is the measurement operator, and the gap $\Delta$ is set as $0.2$.  The label of $\bm{x}_i$ is assigned as $y_i=0$ if 
$$\langle g(\omega_1^{(i)},\omega_2^{(i)})|V^{\dagger}\Pi V|g(\omega_1^{(i)},\omega_2^{(i)}) \rangle\leq 0.5 -\Delta.$$ 
We illustrate the synthetic dataset $\hat{\mathcal{D}}$ in the left panel of Figure \ref{fig:kernel_data}. 

At the data preprocessing stage, we split the dataset $\hat{\mathcal{D}}$ into the training datasets $\hat{\mathcal{D}}_{train}$ with size $N_{train}=100$ and the test dataset $\hat{\mathcal{D}}_{test}$ with $N_{test}=100$.  In the training process, we follow the construction rule of GBLS to build the extended training dataset $\mathcal{D}_{train}$ by using $\hat{\mathcal{D}}_{train}$. We set $K=4$ in the following analysis, where the training example $\mathcal{D}_k\subset \mathcal{D}_{train}$ can be encoded into a quantum state by using four qubits with $N_I=N_F=2$ (see  \ref{appd:simulation} for the detailed implementation of GBLS).  Note that, at each epoch, we shuffle $\mathcal{D}_{train}$ and rebuild the extended dataset $\mathcal{\hat{D}}_{train}$.  An epoch means that an entire dataset is passed forward through the quantum learning model, e.g., when the dataset contains $1000$ training examples, and only two examples are fed into the quantum learning model each time, then it will take $500$ iterations to complete $1$ epoch.

The numerical simulations are implemented on Python in conjunction with the PennyLane, Qiskit, and pyQuil libraries \cite{bergholm2018pennylane,aleksandrowicz2019qiskit,smith2016practical}. The hyper-parameters setting used in our experiment is as follows. The block of $U_E$ in Figure \ref{fig:GBLS_unit_UE} is employed once for the case $K=4$, according to the Grover's theorem $O(\sqrt{K})$. The layer number of variational quantum circuits, i.e., $U_{L_1}=\prod_{l=1}^{L}U(\bm{\theta}^l)$, is set as $L=2$. The number of epochs used in classical optimization is $20$.  For comparison, we also apply the quantum kernel classifier proposed by \cite{havlivcek2019supervised,schuld2019quantum} with two different loss functions, i.e., the mean squared error (MES) loss, and the binary cross entropy (BCE) loss, to learn the synthetic dataset $\mathcal{\hat{D}}$. The selection of the quantum kernel classifiers as the reference is based on the fact that this method has achieved  state-of-the-art performance to classify nonlinear data \cite{havlivcek2019supervised}.  

\textbf{Ideal setting.} We first evaluate performance of different quantum classifiers under the ideal setting, where the quantum system is noiseless and the number of measurements is infinite. The right panel of Figure \ref{fig:kernel_data} illustrates the averaged training and testing accuracies versus the number of epochs. In particular, our proposal achieves comparable performance with the quantum kernel classifier with the BCE loss, where both the train and test accuracies converge to $99\%$ within $2$ epochs. Moreover, these two methods outperform the quantum kernel classifier with the MSE loss ($B=N$), whose test accuracy can only reach $95\%$ after $10$ epochs. The variance of these three quantum classifiers after $10$ epochs becomes small, which implies that all of them hold stable performance under the ideal setting.  

\begin{figure*}[!h]  
\centering
\includegraphics[width=0.955\textwidth]{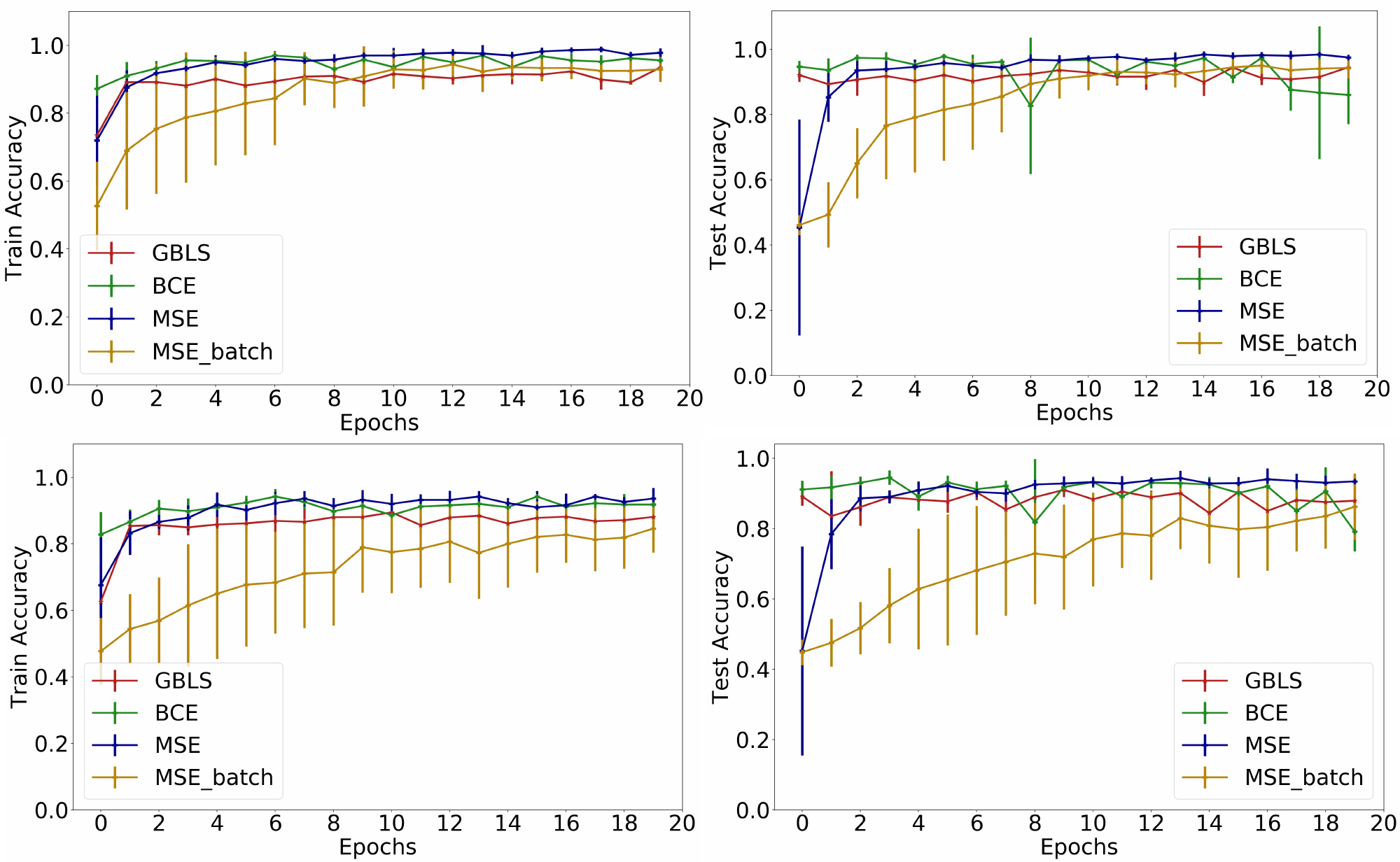}
\caption{\small{\textbf{The performance of different quantum classifiers with finite measurements under the varied depolarization rates.} The labels `GBLS', `BCE', `MSE', and `MSE\_batch' refer to the proposed Grover-based quantum classifier, the quantum kernel classifier with the binary cross entropy loss, the quantum kernel classifier with the mean square error loss and the number of batches being $B=N$, and the quantum kernel classifier with the mean square error loss and the number of batches being $B=N/4$. The upper panel and the lower panel demonstrate the train and test accuracies of GLBS and the quantum kernel classifier with BCE loss when the depolarization rate is set as $p=0.05$ and $p=0.25$, respectively. The vertical sticks reflect the variance of the train and test accuracy at each iteration. }}
\label{fig:kernel_data_noisy}
\end{figure*}

\textbf{Depolarization noise setting.}  We next investigate  performance of GBLS and the referenced quantum kernel classifiers under the realistic setting, where the quantum system noise is considered and the number of measurements is finite. Specifically, we employ the depolarization channel to model the system noise, i.e., given a quantum state $\rho\in \mathbb{C}^{d\times d}$, the quantum depolarization channel $\mathcal{E}_{p}$ that acts on this state is defined as 
$$\mathcal{E}_{p}(\rho) =    (1-p)\rho  + p\pi_{d},$$ 
where $p$ is the depolarization rate, and $\pi_{d}$ is the maximally mixed state with $\pi_{d}   = \mathbb{I}_d/d$. Meanwhile, to explore the trade-off between the computational cost (i.e., the total number of measurements) and the utility $R_1$ indicated by Theorem \ref{thm:informal_utl_QNNQAE_DP}, we also compare  performance between GBLS and a modified quantum kernel classifier with the MSE loss, which supports to use the batch gradient descent method with $B=N/4$ to optimize parameters (Please refer to \ref{appd:simulation} for implementation details). Table \ref{tab:my-table} summarizes the basic information about GBLS and the referenced quantum classifiers.  See \ref{appd:query_comp} about the derivation of the required number of measurements for GBLS and the quantum kernel classifier with the BCE loss.

\begin{center} 
\begin{table}[h!]
\begin{tabular}{|c|c|c|c|c|}
\hline
\centering
Methods & MSE\_batch  & MSE  & BCE  & GBLS   \\ \hline
Number of batches $B$  & $\frac{N}{K}$  & $N$  & $N$  & $\frac{N}{K}$   \\ \hline
 Number of measurements & $O(\frac{TMNd}{K})$ & $O(TMNd)$ & $O(TMNd)$                                     & $O(\frac{TMNd}{K})$  \\ \hline
\end{tabular}
\caption{The basic information of different quantum classifiers. The notations $T$, $K$, $M$, $N$, and $d$ refer to  the number of epochs, the batch size (i.e., in our simulation $K=4$), the number of measurements used to estimate quantum expectation value, the total number of training examples, and the total number of trainable parameters.}
\label{tab:my-table}
\end{table}
\end{center}

The hyper-parameters settings applied to GBLS and other quantum classifiers are as follows. The depolarization rate is set as $p=0.05$ and $p=0.25$, respectively. The number of measurements is set as $10$ to approximate the quantum expectation result. The parameter shift rule is used to estimate the analytic gradients \cite{mitarai2018quantum,schuld2019evaluating}. For each classifier, we repeat the numerical simulations with five times to collect the statistical information. Confer \ref{appd:simulation} for other settings such as learning rates and random seeds.

   The simulation results of GBLS and the referenced quantum classifiers  are illustrated in Figure \ref{fig:kernel_data_noisy}. Specifically, when $p=0.05$, GBLS and the other three referenced quantum classifiers achieve comparable performance after $10$ epochs. Moreover, the quantum kernel classifier with the MSE loss ($B=N/4$ possesses a lower the convergence rate and a larger variance than the rest three classifiers. When $p=0.25$, there exists a relatively large gap between the quantum kernel classifiers with  the MSE\_bactch method and the rest three quantum classifiers in the measure of the convergence rate.  Such a difference reflects the importance to use GBLS to investigate classification tasks under the varied number of batches. We summarize the averaged training and test accuracies of GBLS and other quantum classifiers at the last epoch in Table \ref{tab:depolar_GBLS}.  Even though the measurement error and quantum gate noise are considered, GBLS can still attain stable performance, since its variance is very small (i.e., at most $0.04$). This observation suggests the applicability of our proposal on NISQ machines.     
   
   \begin{center} 
\begin{table}[h!]
\begin{tabular}{|c|c|c|c|c|}
\hline
\centering
Methods & MSE\_batch  & MSE  & BCE  & GBLS   \\ \hline
$p=0.05$ (train)  & $0.929\pm 0.037$  & $0.978 \pm 0.013$  & $0.956 \pm 0.024$  & $0.935\pm 0.024$   \\ \hline
$p=0.25$ (train) & $0.846 \pm 0.072$ & $0.936\pm 0.032$ & $0.918\pm 0.031$                                     & $0.881\pm 0.025$ \\ \hline
$p=0.05$ (test)  & $0.943 \pm 0.032$  & $0.975 \pm 0.006$  & $0.860 \pm 0.089$  & $0.945\pm 0.021$   \\ \hline
$p=0.25$ (test) & $0.862 \pm 0.095$ & $0.934\pm 0.009$ & $0.791\pm 0.056$                                     & $0.879\pm 0.040$ \\ \hline
\end{tabular}
\caption{Performance of different quantum classifiers under the depolarization noise at the $20$-th epoch. The labels `MSE\_batch', `MSE', `BCE', and `GBLS' follow the same meanings as explained in Table \ref{tab:my-table}. The value `$a\pm b$' refers that the averaged accuracy is $a$ and its variance is $b$.   }
\label{tab:depolar_GBLS}
\end{table}
\end{center}

\begin{figure*}[h!]
	\centering
	\includegraphics[width=0.98\textwidth]{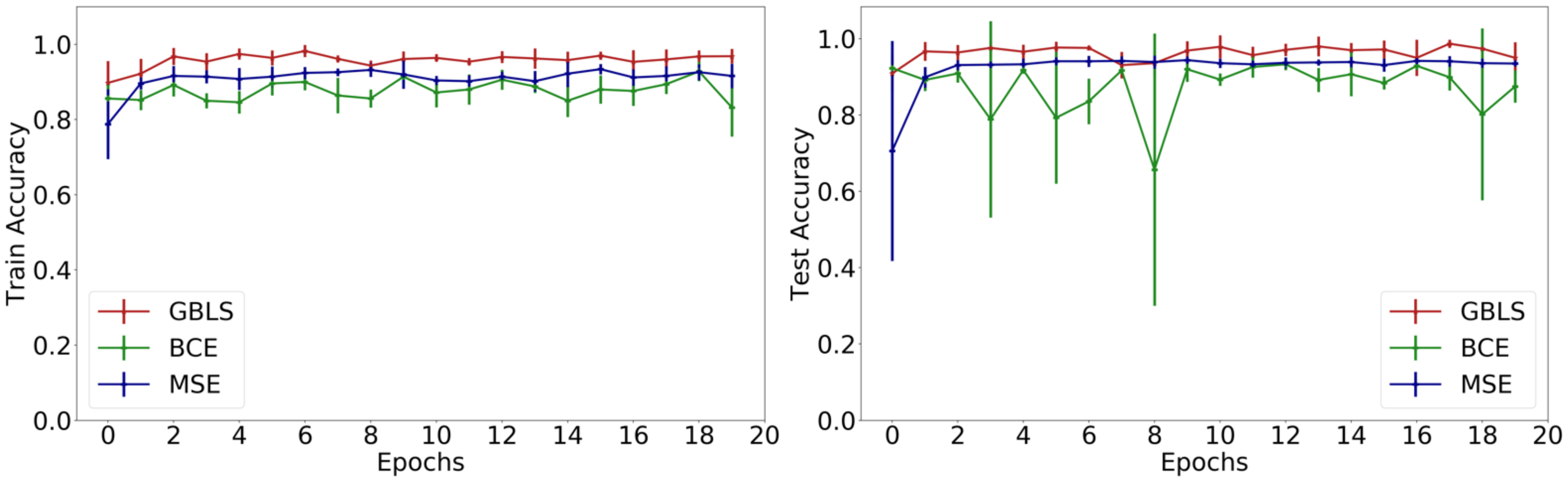}
	\caption{\small{\textbf{Simulation results of different quantum classifiers under the realistic noise setting}.} The labels `GBLS', `BCE', and `MSE' have the same meaning as explained in Figure \ref{fig:kernel_data_noisy}. The noise model, which is extracted from a real quantum hardware, is applied to the trainable unitary $U_L(\bm{\theta})$ of these three classifiers. }
	\label{fig:sim_IBM} 
\end{figure*}

We would like to emphasize the main issue considered in this study: whether there exists a quantum classifier that can attain a good utility bound $R_1$ by using a few number of measurements. The numerical simulation results of GBLS provide a positive response towards this issue. Recall the setting given in Table \ref{tab:my-table} and the results in Figure \ref{fig:kernel_data_noisy}. Although the required number of measurements for GBLS is reduced by $K=4$ times compared with quantum classifiers with the BCE loss and the MSE loss ($B=N$), they achieve comparable performance.  This result implies a huge separation of the computational efficacy between GBLS and previous quantum classifiers with $B=N$  when $N$ is large.

\textbf{Noise model from real quantum hardware.} We further compare performance of GBLS and the referenced quantum classifiers under a noise model extracted from real quantum hardware, i.e., IBMQ\_ourense, provided by the Qiskit and PennyLane Python Libraries \cite{bergholm2018pennylane,aleksandrowicz2019qiskit}. Notably, for all classifiers, the gate noise is only imposed on the trainable quantum circuits $U_L$ instead of the whole circuits, since the implementation of multi-controlled gates (e.g., CCZ) used in GBLS will introduce a huge amount of noise and destroy the optimization of GBLS (See \ref{appd:simulation} for details). Meanwhile, the measurement noise is applied to all quantum classifiers. Due to the relatively poor performance of the quantum kernel classifier with the MSE loss and $B=N/4$, here we only focus the comparison among GBLS and quantum kernel classifiers with the BCE loss and the MSE loss ($B=N$).  Note that all hyper-parameters settings are identical to those used in the above numerical simulations.

The simulation results are exhibited in Figure \ref{fig:sim_IBM}. Specifically, the three classifiers achieve  comparable performance. Such results indicate that the efficacy of GBLS, since the required number of measurements for GBLS is reduced by four times compared with the rest two quantum classifiers. 

\section{Discussion and Conclusion}
In this study, we have proposed a Grover-search based learning scheme for classification. Different from previous proposals, GBLS supports the optimization of a wide range of quantum classifiers with a varied number of batches. This property allows us to explore the trade-off between  the computational efficiency and the utility bound $R_1$.  Moreover, we demonstrate that  GBLS possesses a potential advantage to tackle certain classification tasks in the measure of query complexity. Numerical experiments showed that GBLS can achieve comparable performance  with other advanced quantum classifiers by using a fewer number of measurements. We believe that our work will provide immediate and practical applications for near-term quantum devices.

\medskip

\providecommand{\newblock}{}

 \newpage
\appendix

\section{Proof of Theorem $1$}\label{appd:thm1}
\textbf{Proof of Theorem  $1$.}\\
To achieve Theorem 1, we separately discuss the situations in which the label of the last entry in $\mathcal{D}_k$ is $y_k=1$ and $y_k=0$, respectively.  
 
\noindent\textit{\underline{For the case $y_k=1$.}} Suppose that the label of the last entry in $\mathcal{D}_k$ is $y_k=1$. Followed from Eqn.~(\ref{eqn:init}), after the first cycle, the generated state of GBLS is 
\begin{equation}\label{eqn:prf1}
U\ket{\bm{0}}_{F,I}\equiv U_{c_1}\ket{\Phi^k(y_k=1)}_{F,I}= \ket{\bm{0}}_F\otimes\left(\cos 3\gamma\ket{B}_I + \sin3\gamma\ket{i^*}_I  \right)\nonumber,
\end{equation}
 where $ \sin\gamma=\frac{1}{\sqrt{K}}$. This result indicates that the probability to sample the target index $i^*$ is increased from $\sin^2 \gamma$ to $\sin^23\gamma$, which is same with Grover-search.  

Then, by induction as the proof of Grover-search does \cite{brassard2000quantum}, the generated state of GBLS after applying $U$ to $\ket{\bm{0}}_{F,I}$ with $\ell$ times yields 
\begin{equation}\label{eqn:prf5}
\prod_{i=1}^{\ell}U^i\ket{\bm{0}}_{F,I}=\ket{\bm{0}}_F\otimes(\cos ((2\ell+1)\gamma)\ket{B}_I +\sin((2\ell+1)\gamma) \ket{i^*}_I). 
\end{equation}

Note that, GBLS requires that the employed quantum operation at the last cycle is  $U_E$ as defined in Eqn.~(\ref{eqn:U_E}) instead of $U$. Mathematically, the generated state is
\begin{align}\label{eqn:thm1_eqn1}
	U_E\prod_{i=1}^{\ell}U^i\ket{\bm{0}}_{F,I} & =U_{init}\circ \MCZ  \circ (U_{L_1}\otimes \mathbb{I})\circ U_{data}\ket{\bm{0}}_F\otimes(\cos ((2\ell+1)\gamma)\ket{B}_I +\sin((2\ell+1)\gamma) \ket{i^*}_I) \nonumber\\
	& = U_{init}\circ \MCZ \left(\cos ((2\ell+1)\gamma)\ket{\psi_B^{(0)}}_F\ket{B}_I +\sin((2\ell+1)\gamma) \ket{\psi_B^{(1)}}_F\ket{i^*}_I) \right) \nonumber\\
	& = U_{init} \left(\cos ((2\ell+1)\gamma)\ket{\psi_B^{(0)}}_F\ket{B}_I - \sin((2\ell+1)\gamma) \ket{\psi_B^{(1)}}_F\ket{i^*}_I) \right) \nonumber\\ 
	& = \left(\cos ((2\ell+3)\gamma)\ket{\psi_B^{(0)}}_F\ket{B}_I + \sin((2\ell+3)\gamma) \ket{\psi_B^{(1)}}_F\ket{i^*}_I) \right)~,
\end{align}
where the first equality uses Eqn.~(\ref{eqn:prf5}), the second equality exploits Eqn.~(\ref{eqn:init}) to engineer the feature register, the third equality employs $\MCZ$ to flip the phase the state $\ket{i^{*}}$ whose first qubit in the feature register is $\ket{1}$, and last equality comes from the application of the diffusion operator $U_{init}=\mathbb{I}_{F}\otimes  (2\ket{\varphi}\bra{\varphi} - \mathbb{I}_I)$  with $\ket{\varphi}=\frac{1}{\sqrt{K}}\sum_i \ket{i}$ to the index register.    

The result of Eqn.~(\ref{eqn:thm1_eqn1}) indicates that, under the optimal setting, the probability to sample $i^*$ is close to $1$ when $\ell\sim O(\sqrt{K})$, since $\sin \gamma\approx \gamma=1/\sqrt{K}$ and then  $\sin ((2\ell+3)\gamma)\approx 1$.   

\noindent\textit{\underline{For the case $y_k=0$.}} We then demonstrate that, when the label of the last entry in $\mathcal{D}_k$ is $y_k=0$, even if applying $U=\prod_{i=1}^{\ell}$ and $U_E$ to $\ket{\bm{0}}_{F,I}$ with $\ell\sim O(\sqrt{K})$, the probability  to sample $i^*$ is $1/K$. Followed from Eqn.~(\ref{eqn:opt_lab0}), after the first cycle, the generated state of GBLS is 
\begin{equation}\label{eqn:prf1}
U_{c_1}\ket{\Phi^k(y_k=0)}_{F,I}=  \frac{1}{\sqrt{K}}\sum_{i=0}^{K-1}\ket{\bm{0}}_F\ket{i}_I\nonumber~,
\end{equation}
where $\sin\gamma=\frac{1}{\sqrt{K}}$. Due to $U_{c_1}\ket{\Phi^k(y_k=0)}_{F,I}=U\ket{\bm{0}}_{F,I}$,   after applying $U$ to the state $\ket{\bm{0}}$, the probability to sample any index is identical. By induction, applying the corresponding $U$ to the state $\ket{\bm{0}}_{F,I}$ with $\ell$ times yields
 \begin{equation}\label{eqn:prf2}
\prod_{i=1}^{\ell}U^i\ket{\bm{0}}_{F,I}= \frac{1}{\sqrt{K}}\sum_{i=0}^{K-1}\ket{\bm{0}}_F\ket{i}_I, 
\end{equation}
where given any positive integer $\ell$, the probability to sample $\ket{i^*}_{I}$ is $1/K$. 

As with the case of $y_k=1$, at the last cycle, we apply the unitary $U_E$ to the state $\prod_{i=1}^{\ell}U^i\ket{\bm{0}}_{F,I}$, and the generated state is
\begin{align}\label{eqn:prf3}
\allowdisplaybreaks
	U_E\prod_{i=1}^{\ell}U^i\ket{\bm{0}}_{F,I} &= U_{init}\circ \MCZ  \circ (U_{L_1}\otimes \mathbb{I})\circ U_{data}\frac{1}{\sqrt{K}}\sum_{i=0}^{K-1}\ket{\bm{0}}_F\ket{i}_I \nonumber\\
	&=  U_{init}\left( \frac{1}{\sqrt{K}}\sum_{i=0}^{K-1} \left(\ket{\psi_B^{(0)}}_F\ket{B}_I +  \ket{\psi_{i^*}^{(0)}}_F\ket{i^{*}}_I\right) \right) \nonumber\\
	&= \frac{1}{\sqrt{K}}\sum_{i=0}^{K-1} \left(\ket{\psi_B^{(0)}}_F\ket{B}_I +  \ket{\psi_{i^*}^{(0)}}_F\ket{i^{*}}_I\right)~,
\end{align}
where the first equality uses the explicit form of $U_E$ and Eqn.~(\ref{eqn:prf2}), and the second equality is guaranteed by Eqn.~(\ref{eqn:opt_lab1}) (note that the only difference is replacing $\ket{\psi_{i^*}^{(1)}}_F$ with $\ket{\psi_{i^*}^{(0)}}_F$ based on the setting $y_k=0$), and the last equality exploits the explicit form of $U_{init}$. 
  
The result of Eqn.~(\ref{eqn:prf3}) reflects that, under the optimal setting, the probability to sample $i^*$ can never be increased when $y_k=0$. Therefore, we can conclude that, under the optimal setting, the probability to sampling the outcome $i^*$  approaches $1$ asymptotically if and only if the label of the last entry of $\mathcal{D}_k$ is $y_k=1$.   $\square$

\section{Variational quantum circuits and the optimizing method}\label{appd:PQC}

In this section, we first introduce the variational quantum circuits $U_{L_1}(\bm{\theta})$ used in GBLS. We then elaborate the optimization method, i.e., the parameter shift rule, that is employed to train $U_{L_1}(\bm{\theta})$. 

\begin{figure}[h!] 
\centering
\includegraphics[width=0.75\textwidth]{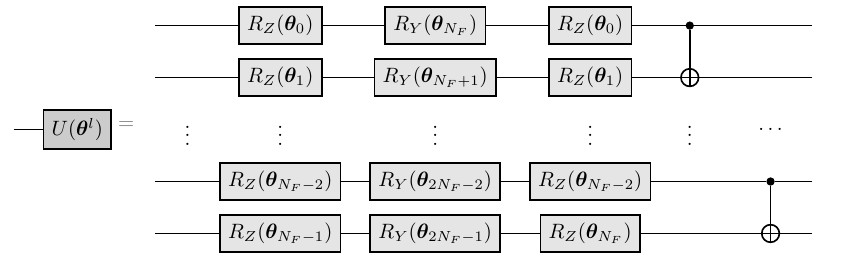}
\caption{The implementation of the $l$-th layer $U(\bm{\theta}^l)$. Suppose that the $l$-th layer $U(\bm{\theta}^l)$ interacts with $N_F$ qubits. Three trainable parameterized gates, $R_Z$, $R_Y$ and $R_Z$, are firstly applied to each qubit, followed by $N_F - 1$ CNOT gates.} 
\label{fig:MPQC}
\end{figure}

Variational quantum circuits, which is also called parameterized quantum circuit, are composed of trainable single qubit gates and two qubits gates (e.g., CNOT or CZ). As a promising scheme for NISQ devices, variational quantum circuits have been extensively investigated for accomplishing the generative  and discriminative \cite{benedetti2019generative,du2018expressive,farhi2018classification,schuld2020circuit,grant2018hierarchical} tasks via variational hybrid quantum-classical algorithms \cite{mcclean2016theory}. One typical variational quantum circuits is called the  multiple-layer parameterized quantum circuits (MPQC), where the arrangement of quantum gates in each layer is identical \cite{benedetti2019generative}. Denote the operation formed by the $l$-th layer as $U(\bm{\theta}^l)$. The generated quantum state from MPQC yields   \[\ket{\Psi} =  \prod_{l=1}^{L}U(\bm{\theta}^l)\ket{0}^{\otimes N_F}~,\] where  $L$ is the total number of layers. GBLS employs  MPQC to construct $U_{L_1}$, i.e., 
\begin{equation}\label{eqn:appA_pqc}
	U_{L_1}(\bm{\theta}) = \prod_{l=1}^{L}U(\bm{\theta}^l)~,
\end{equation}
  and the circuit arrangement for the $l$-th layer $U(\bm{\theta}^l)$ is shown in Figure \ref{fig:MPQC}. When the number of layers is $L$, the total number of trainable parameters for GBLS is $2N_FL$. 
				
The updating rule of GBLS at the $k$-th iteration follows  
\begin{equation}\label{eqn:appd_update_rule0}
	\bm{\theta}^{(k+1)} = \bm{\theta}^{(k)} - \eta \frac{\mathcal{L}(\bm{\theta}^{(k)}, \mathcal{D}_k)}{\partial\bm{\theta}}~,
\end{equation}
where $\eta$ is the learning rate and $\mathcal{D}_k$ is the $k$-th training example. By expanding the explicit form of $\mathcal{L}(\bm{\theta}^{(k)}, \mathcal{D}_k)$ given in Eqn.~(\ref{eqn:loss}), the gradients of $\mathcal{L}(\bm{\theta}^{(k)}, \mathcal{D}_k)$ can be rewritten as  
\begin{equation}\label{eqn:appd_update_rule01}
	\frac{\partial \mathcal{L}(\bm{\theta}^{(k)}, \mathcal{D}_k)}{\partial\bm{\theta}} = \text{sign}(1/2-y_k)\frac{\partial \Tr(\Pi \rho(\bm{\theta}^{(k)}))}{\partial \bm{\theta}}~,
\end{equation}
where $y_k$ refers to the label of the last entry in $\mathcal{D}_k$, $\text{sign}(\cdot)$ is the sign function, $\Pi$ is the measurement operator, and 
\[ \rho(\bm{\theta}^{(k)})=U_EU(\bm{\theta}^{(k)})^{O(\sqrt{K})}\ket{\bm{0}}\bra{\bm{0}}(U_EU(\bm{\theta}^{(k)})^{O(\sqrt{K})})^{\dagger}~.\] 
GBLS adopts the parameter shift rule proposed by \cite{mitarai2018quantum} to  attain  the gradient $\frac{\partial \Tr(\Pi \rho(\bm{\theta}^{(k)}))}{\partial \bm{\theta}}$. Concisely, the parameter shift rule iteratively computes each entry of the gradient.   Without loss of generality, here we explain how to compute $\frac{\partial \Tr(\Pi \rho(\bm{\theta}^{(k)}))}{\partial \bm{\theta}_j}$ for $j\in[2N_FL]$. Define  $\bm{\theta}^{(k)}_{\pm}$ as 
\begin{equation}\label{eqn:appA_theta}
	\bm{\theta}^{(k)}_{\pm} = [\bm{\theta}^{(k)}_0,...,\bm{\theta}^{(k)}_{j-1}, \bm{\theta}^{(k)}_{j}\pm \frac{\pi}{2},\bm{\theta}^{(k)}_{j+1},...,\bm{\theta}^{(k)}_{2N_FL-1}]~,
\end{equation}
 where only the $j$-th parameter is rotated by $\pm \frac{\pi}{2}$. Then the mathematical representation of the  gradient for the $j$-th entry is 
 \begin{equation}\label{eqn:appd_update_rule1}
 	\frac{\partial \Tr(\Pi \rho(\bm{\theta}^{(k)}))}{\partial \bm{\theta}_j} = \frac{\Tr(\Pi \rho(\bm{\theta}^{(k)}_+))- \Tr(\Pi \rho(\bm{\theta}^{(k)}_-))}{2}~.
 \end{equation}

 In conjunction with Eqn.~(\ref{eqn:appd_update_rule0}), Eqn.~(\ref{eqn:appd_update_rule01}), and Eqn.~(\ref{eqn:appd_update_rule1}), the updating rule of GBLS at the $t$-th iteration for the $j$-th entry is
 \begin{equation}\label{eqn:grad_GBLS}
 		\bm{\theta}^{(k+1)}_j = \bm{\theta}^{(k)}_j - \eta  \frac{\Tr(\Pi \rho(\bm{\theta}^{(k)}_+))- \Tr(\Pi \rho(\bm{\theta}^{(k)}_-))}{2} \text{sign}(\frac{1}{2}-y_k)~.
 \end{equation}

\begin{figure*}   
\centering 
\includegraphics[width=0.65\textwidth]{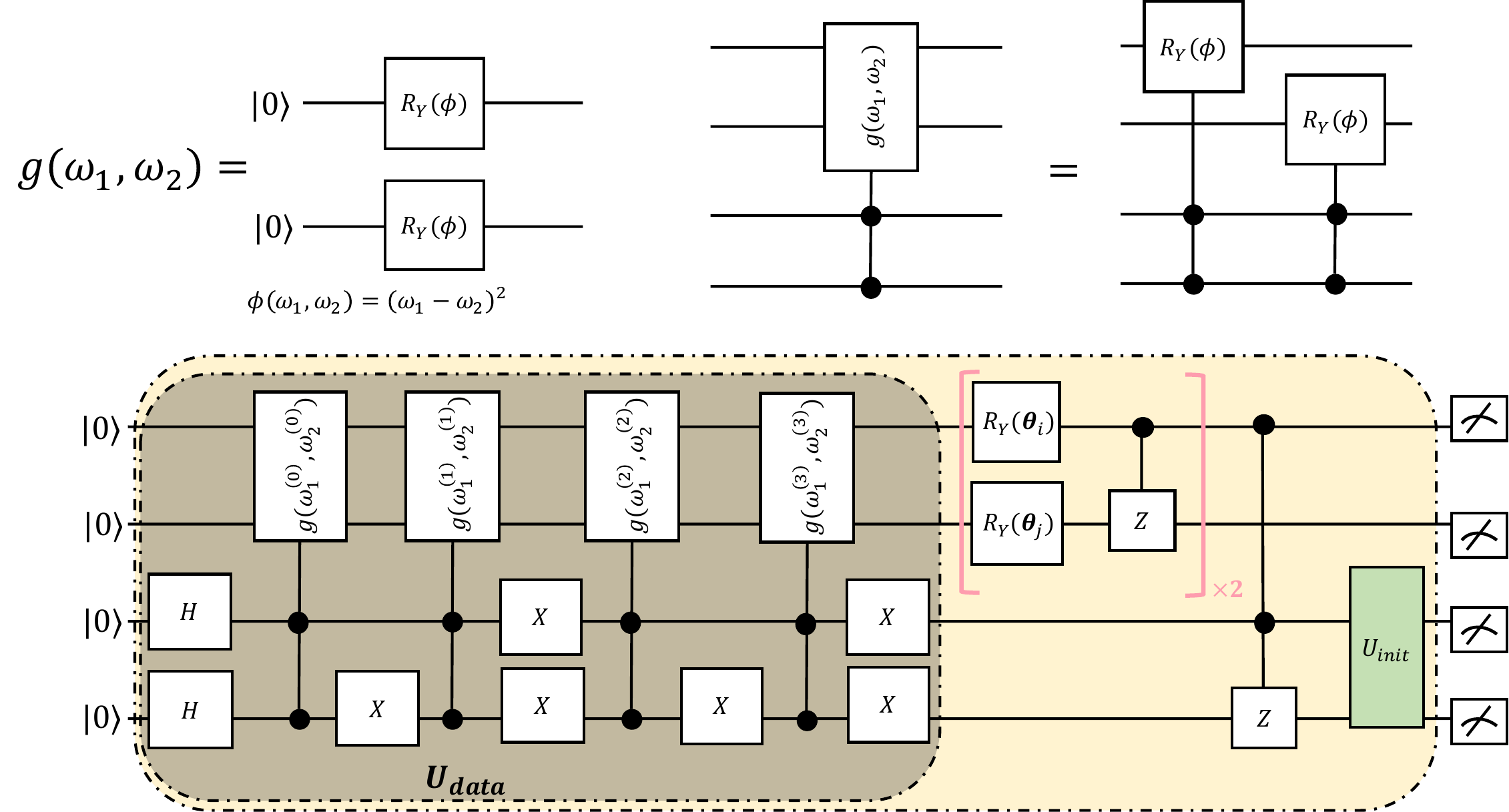}
\caption{The implementation of GBLS used in numerical simulations. The upper left panel illustrates the circuit implementation of the encoding unitary $U_{data}$ corresponding to the feature map  $g(\bm{x}_i)$. The lower panel demonstrates the implementation of GBLS given the input $\mathcal{D}_k=\{\bm{x}_i,\bm{x}_j,\bm{x}_k,\bm{x}_l\}$, where the implementation of the controlled-$g(\bm{x}_i)$ quantum operation is shown in the upper right panel. }
\label{fig:kernel_circuit}
\end{figure*}	

\section{More details of numerical simulations}\label{appd:simulation}
In this section, we provide more  details about the numerical simulations. Specifically, we first explain how to construct the employed synthetic dataset. We then elaborate on the implementation of GBLS and referenced classifiers, and their hyper-parameters settings. We next analyze the required circuit depth to implement these quantum classifiers. Last, we introduce the construction of the modified dataset used in the MSE\_batch method.

\textbf{The construction of the synthetic dataset.} Given the training example $\bm{x}_i=(\omega_1^{(i)}, \omega_2^{(i)})\in\mathbb{R}^2$ for all $i\in[N-1]$, the embedding function $g(\omega_1^{(i)}, \omega_2^{(i)}):\mathbb{R}^2\rightarrow\mathbb{C}^4$ that is used to encode $\bm{x}_i$ into the quantum states is formulated as 
\begin{equation}\label{eqn:appC_encode}
g(\omega_1^{(i)}, \omega_2^{(i)})=\left(R_Y(\phi(\omega_1^{(i)}, \omega_2^{(i)}))\otimes R_Y(\phi(\omega_1^{(i)}, \omega_2^{(i)}))\right) \ket{0}^{\otimes 2} ~,
\end{equation} 
where  $\phi(\omega_1^{(i)}, \omega_2^{(i)})=(\omega_1^{(i)}-\omega_2^{(i)})^2$ is a specified mapping function.  The above formulation implies that  $g(\bm{x}_{i})$ can be converted to a sequence of quantum operations, where its implementation is illustrated in the upper left panel of Figure  \ref{fig:kernel_circuit}. To simultaneously encode multiple training examples into the quantum states, we should implement $g(\bm{x}_i)$ as a controlled version, where the implementation is shown in the upper right panel of Figure  \ref{fig:kernel_circuit}.
 
The random unitary $V\in SU(4)$ used in the numerical simulations is formulated as $V=R_Y(\psi_1)\otimes R_Y(\psi_2)$, where $\psi_1$ and $\psi_1$ are uniformly sampled from $[0, 2\pi)$. 

\textbf{The details of GBLS, the referenced classifiers, and hyper-parameters setting.} The implementation of GBLS is shown the lower panel of Figure  \ref{fig:kernel_circuit}. In particular, the data encoding unitary $U_{data}$ is composed of a set of controlled-$g(\bm{x}_i)$ quantum operations. The MPQC introduced in \ref{appd:PQC} is employed to build $U_{L_1}(\bm{\theta})$, where each layer $U(\bm{\theta}^l)$ is composed of $R_Y$ gates and CZ gates and the layer number is $L=2$. 

The basic components of the referenced quantum classifiers are identical to those used in GBLS. In particular, for all employed quantum kernel classifiers, the implementation of variational quantum circuits $U_{L_1}(\bm{\theta})$ are the same with GBLS, where the layer number is $L=2$ and each layer is composed of $R_Y$ gates and CZ gates as shown in Figure \ref{fig:kernel_circuit}. The implementation of the encoding unitary $U_{data}$ depends on the batch size $B$. For the quantum kernel classifiers with the BCE loss and MSE loss ($B=N$), following Eqn.~(\ref{eqn:appC_encode}), the encoding unitary is 
\begin{equation}\label{eqn:u_data_b=1}
	U_{data}=R_Y(\phi(\omega_1^{(i)}, \omega_2^{(i)}))\otimes R_Y(\phi(\omega_1^{(i)}, \omega_2^{(i)})).
\end{equation}  
For the quantum kernel classifier with the MSE loss ($B=N/4$), the implementation of the encoding unitary $U_{data}$ is the same with  GBLS as shown in Figure \ref{fig:kernel_circuit}.

The detailed hyper-parameters settings for GBLS and the referenced classifiers are as follows. The learning rate for GBLS, the quantum kernel classifier with the BCE loss, the quantum kernel classifier with the MSE loss ($B=N$ and $B=N/4$) is identical, which is set as $\eta = 1.0$. Moreover, when we explore the statistical performance of different quantum classifiers under the noise setting, the random seeds are set as $\{i\}_{i=1}^R$ with $R$ being the total number of repetitions.

\textbf{The analysis of the quantum circuit depth.} 
Here we analyze the required circuit depth to implement quantum kernel classifiers used in numerical simulations.  As explained in the above subsection, the quantum kernel classifiers with $B=N$ can be efficiently realized, since the data encoding unitary $U_{data}$ and the variational quantum circuits only involve single and two qubits gates. In particular, the circuit depth to construct the unitary $U_{data}$ in Eqn.~(\ref{eqn:u_data_b=1}) is $1$. Moreover, the circuit depth to construct $U_L(\bm{\theta})$ as shown in Figure \ref{fig:kernel_circuit} is $4$. In total, when the number of batches $B$ equals to $N$, the required depth for the quantum kernel classifier  with the BCE or MSE  loss is $5$.

Compared with the setting $B=N$, the implementation of the quantum kernel  classifier with $B=N/4$ and GBLS requires a relatively deep circuits. The substantial reason is that the fabrication of the data encoding unitary $U_{data}$ involves multi-controlled qubits gates as shown in Figure \ref{fig:kernel_circuit} (highlighted by the brown region). Specifically, when we decompose the CC-$R_Y$ gate into single-qubit and two-qubit gates, the required circuit depth is $27$. Therefore, following Figure \ref{fig:kernel_circuit}, the circuit depth to implement $U_{data}$ is $113$. Considering that the circuit depth to implement $U_{L_1}$ is $4$, the total circuit depth to implement the quantum kernel classifier with $B=N/4$ is $117$. As shown in Figure \ref{fig:kernel_circuit},  the quantum circuit in GBLS is composed of $U_{data}$, $U_{L_1}$, and $U_{init}$. The implementation of $U_{data}$ and $U_{L_1}$ is identical to the quantum kernel  classifier with $B=N/4$. Moreover, based on Grover-search algorithm, the circuit depth to implement $U_{init}$ is $15$, which includes  $4$ Hadamard gates and $1$ CCZ gate. Therefore, the total circuit depth to implement GBLS is $132$. 

We remark that the circuit depth of the quantum kernel  classifier with $B=N/4$ and GBLS is dominated by the implementation of $U_{data}$, which exploits multi-controlled qubits gates to load different training  examples in superposition. Such an observation implies that efficient encoding methods can dramatically reduce the required circuit depth to construct these quantum classifiers. A possible solution is proposed by \cite{heya2018variational}, which constructs a target multi-qubits gate by optimizing a variational quantum circuit which consists of tunable single-qubit gates and fixed two qubits gates. 

\textbf{The modified training dataset for the MSE\_batch method.} We note that naively employing the original training dataset $\hat{\mathcal{D}}$ to optimize the quantum kernel classifier with the MSE\_batch loss is infeasible. Let us illustrate a simple example. Suppose the input state is $\frac{1}{\sqrt{2}} \sum_{i=1}^2 \ket{g(\bm{x}^{(i))}}_F\ket{i}_I$ with the batch size $2$, where the subscript `I' (`F') refers to the index (feature) register. When   the trainable quantum circuits $U_L(\bm{\theta})\otimes \mathbb{I}_I$ and  the measurement operator are applied to this state, the output corresponds to the averaged predictions of the examples $\{\bm{x}^{(i))}\}_{i=1}^2$. Such a setting is ill-posed once the labels $\bm{x}^{(1)}$ and $\bm{x}^{(i)}$ of are opposite, e.g., the former is $0$ and the latter is $1$, since a wrong prediction (the former is $1$ and the latter is $0$) also leads to the averaged truth label $0.5$.

To conquer the above issue, we build a modified dataset instead of $\mathcal{\hat{D}}$ to optimize the quantum kernel classifier with the MSE\_batch loss. Specifically, we shuffle the given dataset  $\mathcal{\hat{D}}$ and ensure that for the modified dataset, the training examples in each batch $\mathcal{B}_i$ for $\forall i \in [B]$ must possess the same label. In doing so, the averaged truth label can either be $0$ and $1$ without any confusion. 

\section{The computational complexity of GBLS and the quantum kernel classifier with the BCE loss}\label{appd:query_comp}

We now separately derive the required  number of measurements, or equivalently, the computational complexity, for GBLS and the quantum kernel classifier with the BCE loss at each epoch. For both methods,  the  hyper-parameters setting is supposed to be identical, i.e., the size of the dataset $\mathcal{\hat{D}}$ is $N$, the layer number of MPQC $U_{L_1}$ is $L$, the number of qubits to load data features is $N_F$,  the total number of trainable parameters $\bm{\theta}$ is $N_FL$, and the number of measurements applied to estimate the quantum expectation value is $M$. 

We say one query when the variational quantum circuit used in the quantum classifier takes the encoded data and then be measured by the measurement operator once. Following the training mechanism of the quantum classifier, its query complexity amounts to counting the total number of measurements to the variational quantum circuits to acquire the gradients in one epoch.

We now derive the required number of measurements of the quantum kernel classifier with the BCE loss in one epoch. Given the dataset $\mathcal{\hat{D}}$,  the binary cross entropy loss yields 
\begin{equation}
	\mathcal{L}_{BCE} = -\frac{1}{N}\sum_{i=0}^{N-1}y_i\log(p(y_i))+(1-y_i)\log(1-p(y_i))	~,
\end{equation}
 where $y_i$ is the label of the $i$-th example and $p(y_i)$ is the predicted probability of the label $y_i$, or equivalently,   the output of the quantum circuit used in the quantum kernel classifier
\begin{equation}
	p(y_i) = \Tr(\Pi \rho(\bm{\theta}))~,
\end{equation}
where $\rho(\bm{\theta})=U_{L_1}(\bm{\theta})\ket{g(\bm{x}_i)}\bra{g(\bm{x}_i)}U_{L_1}(\bm{\theta})^{\dagger}$, $U_{L_1}(\bm{\theta})$ refers to variational quantum circuits defined in Eqn.~(\ref{eqn:appA_pqc}),  $\ket{g(\bm{x}_i)}$ represents the encoded quantum state defined in Eqn.~(\ref{eqn:appC_encode}), and $\Pi$ is the measurement operator. Following the parameter shift rule, the derivative of BCE loss satisfies
\begin{equation}
	\frac{\partial \mathcal{L}_{BCE}}{\partial \bm{\theta}_j} = \frac{1}{N}\sum_{i=0}^{N-1} \left(\frac{1 - y_i}{1 - p(y_i)} -\frac{y_i}{p(y_i)} \right)\frac{\Tr(\Pi \rho(\bm{\theta}_+))-\Tr(\Pi \rho(\bm{\theta}_-))}{2}~,
\end{equation}
where $\bm{\theta}_{\pm}$ is defined in Eqn.~(\ref{eqn:appA_theta}). The above equation implies that to acquire the gradients of the BCE loss,  it necessitates to feed the training example one by one to the quantum kernel classifier to estimate $p(y_i)$, and then conduct the classical post-processing to compute the coefficient $\frac{1 - y_i}{1 - p(y_i)} -\frac{y_i}{p(y_i)}$. In other words, the number of batches for this quantum classifier can only be $B=N$.  Since the estimation of $p(y_i)$, $\Tr(\Pi \rho(\bm{\theta}_+))$, and $\Tr(\Pi \rho(\bm{\theta}_-))$ are completed by using $M$ measurements, the derivative $\partial \mathcal{L}_{BCE}/\partial \bm{\theta}_j$ can be estimated by using $3NM$ measurements. Considering that there are in total $N_FL$ trainable parameters, the total number of measurements at each epoch for the quantum kernel classifier with the BCE loss is $3NMN_FL$.  

 Unlike the quantum kernel classifier with the BCE loss,  GBLS uses a simple loss function $\mathcal{L}$ defined in Eqn.~(\ref{eqn:loss}), which  allows us to efficiently acquire the gradient $\partial \mathcal{L}/\partial  \bm{\theta}_j$ by leveraging the superposition property. Recall Eqn.~(\ref{eqn:grad_GBLS}).  The gradient of GBLS  satisfies \[\frac{\partial \mathcal{L}(\bm{\theta}, \mathcal{D}_k)}{\partial \bm{\theta}_j}= \frac{\Tr(\Pi \rho(\bm{\theta}^{(k)}_+)) - \Tr(\Pi \rho(\bm{\theta}^{(k)}_-))}{2} \text{sign}(\frac{1}{2}-y_k)~,\]
where $y_k$ refers to the label of the last pair in the extended training example $\mathcal{D}_k$. The above equation indicates that the gradient for $\mathcal{D}_k$, which contains $K$ training examples in $\mathcal{\hat{D}}$, can be estimated by using $2M$ measurements, where the first (last) $M$ measurements aim to approximate $\Tr(\Pi \rho(\bm{\theta}^{(k)}_-))$ ( $\Tr(\Pi \rho(\bm{\theta}^{(k)}_+))$). Therefore, the total number of measurements to collect $\{ \frac{\partial \mathcal{L}(\bm{\theta}, \mathcal{D}_k)}{\partial \bm{\theta}_j} \}$ for all possible $\mathcal{D}_k$ is $2MB = 2 M N /K$. Considering that there are in total $N_FL$ trainable parameters, the query complexity at each epoch for GBLS is $2N_FLMN/K$. Note that when $K\rightarrow N$, the required number of measurements of GBLS can be dramatically reduced.

To ease of understanding, let us illustrate an intuitive example. Define two extended training examples, where the first one includes all positive examples in $\mathcal{D}$  and one negative example, and the second one includes all negative examples in $\mathcal{D}$  and one positive example. Since these two extended examples cover the whole dataset $\mathcal{D}$,  when GBLS uses these two examples to update $\bm{\theta}$, it completes one epoch. Celebrated by the simple form of $\mathcal{L}$,  the number of measurements to estimate the gradients for the $j$-th entry $\bm{\theta}_j$ given these two extended examples is $O(1)$.  Considering there are in total $O(N_FL)$ trainable parameters, the total number of measurements at each epoch for GBLS is $O(LN_F)$.  

\end{document}